# Giant and Broadband Circular Dichroism from Particle-Hole Symmetry Breaking in Weyl Semimetals

Xiangyu Jiang[1†], Zeping Shi[1†], Yuhan Du[1†], Haonan Chen[2†], Jiayu Wang[2], Wenbin Wu[1], Guangyi Wang[1], Congming Hao[1], Mingfan Yao[1], Mingsen Zhou[1], Xin Chen[1], Chenyao Xu[1], Zhongbo Yan[3], Cheng Zhang[2*], Hai-Zhou Lu[4], Junhao Chu[5,6], Xiang Yuan[1,6*]

[1]State Key Laboratory of Precision Spectroscopy, East China Normal University, Shanghai 200241, China
[2]State Key Laboratory of Surface Physics and Institute for Nanoelectronic Devices and Quantum Computing, Fudan University, Shanghai 200433, China
[3]Guangdong Provincial Key Laboratory of Magnetoelectric Physics and Devices, State Key Laboratory of Optoelectronic Materials and Technologies, School of Physics, Sun Yat-sen University, Guangzhou 510275, China
[4]Shenzhen Institute for Quantum Science and Engineering and Department of Physics, Southern University of Science and Technology (SUSTech), 518055 Shenzhen, China.
[5]Institute of Optoelectronics, Fudan University, 200438, Shanghai, China
[6]School of Physics and Electronic Science, Key Laboratory of Polar Materials and Devices, Ministry of Education, East China Normal University, Shanghai 200241, China

[†]These authors contributed equally to this work.
[*]Correspondence and requests for materials should be addressed to X. Y. (E-mail: xyuan@lps.ecnu.edu.cn) and C. Z. (zhangcheng@fudan.edu.cn)

**Giant and Broadband Circular Dichroism**
**from Particle-Hole Symmetry Breaking in Weyl Semimetals**

## Abstract

Circular dichroism originates from symmetry breaking of material structure, leading to differential absorption of left- and right-circularly polarized light. However, circular dichroism in most materials is inherently weak and spectrally narrow, especially in the mid-to-far infrared. Here, we uncover giant infrared circular dichroism in the magnetic-field-forced Weyl semimetal $Mn(Bi,Sb)_2Te_4$, driven by extreme particle-hole symmetry breaking. Helicity-resolved magneto-infrared spectroscopy reveals circular dichroism exceeding 3000 mdeg (~130 mdeg/nm) with above-degree response extending over the 6-13 μm spectral range. The optical resonances are enhanced by a strong band nesting effect intrinsic to the Landau levels of type-II Weyl dispersion. A symmetry-based $k \cdot p$ model reproduces these magneto-infrared responses and demonstrates that magnetization-induced asymmetric spin-orbit coupling generates particle-hole symmetry breaking, suppressing spin-up, parity-even wavefunction components in the valence Landau band and thereby producing pronounced optical helicity selectivity. Our findings establish particle-hole symmetry breaking as an effective route toward helicity-resolved optical control in quantum materials.

## Introduction

Chirality, a fundamental attribute of matter, profoundly shapes the optical and electronic properties of materials. Interactions between chiral media and circularly polarized light drive applications ranging from structural identification and molecular spectroscopy to emerging optoelectronic technologies[1,2]. Circular dichroism (CD), defined by[3–6] the difference in absorbance between left- and right-handed circularly polarized light, directly reflects material chirality and reveals the nature of symmetry breaking.

Realizing strong CD requires that one circular-polarization channel be strongly suppressed within the interested spectral range, which demands a substantial degree of symmetry breaking. In practice, however, such symmetry breaking (e.g., parity symmetry $\mathcal{P}$ breaking in chiral crystals and time-reversal symmetry $\mathcal{T}$ breaking in magnetic systems) is usually weak[7], resulting in small (typically on the order of 1 mdeg/nm) and spectrally narrow CD, especially in the mid-to-far infrared.

Although Weyl semimetals are naturally expected to host pronounced CD[8] owing to their intrinsic chirality, a single Weyl cone exhibits zero net CD in theory[9,10], cancelled by opposite response from different momentum. When applying a strong magnetic field that breaks $\mathcal{T}$ symmetry, inter-Landau-band–transitions are individually circularly polarized, but still cancelled with each other with opposite angular momentum.

Here, we report giant and broadband infrared CD in the Landau-quantized Weyl semimetal $Mn(Bi_{1-x}Sb_x)_2Te_4$, originating from extreme particle-hole symmetry breaking. Magnetic-field-induced ferromagnetism drives the system into a Weyl semimetal phase where asymmetric spin-orbit coupling forced by $\mathcal{T}$-breaking leads to sharply unequal Fermi velocities between particle and hole bands, differing up to a factor of 16, which fundamentally reshapes the Landau band wavefunctions. Magneto-infrared spectroscopy reveals a series of pronounced inter-band Landau-level transitions, originating from the band nesting effect intrinsic to type-II Weyl fermions. Strikingly, helicity-resolved magneto-infrared spectroscopy reveals that the Weyl fermion couples exclusively to $\sigma_+$ photon in broad spectral range, while $\sigma_-$ channel is strongly suppressed. Our symmetry-based k·p model reproduces the band evolution, Landau-level spectrum, and helicity-selective response. The combined theoretical and experimental results establish particle-hole symmetry breaking as a new route for engineering topological phase and chiral light-matter interactions in quantum materials.

We focus on bulk $MnBi_2Te_4$, which is an A-type antiferromagnetic topological insulator in its ground state. When driven into the ferromagnetic (FM-z) phase by an external magnetic field, the system undergoes a topological phase transition into a type-II Weyl semimetal. In this phase, it hosts a single pair of Weyl nodes related by inversion symmetry along the Γ-Z direction[11–13]. Despite fermiology[13,14], this ideal Weyl phase in the bulk FM state has remained largely unexplored. Importantly, while $\mathcal{P}$ is preserved, the broken $\mathcal{PT}$ symmetry in the FM-z state in principle allows non-zero CD[15,16]. (Extended Data Fig. 1)

**Weyl Nodes and Particle-Hole Asymmetry in $Mn(Bi_{1-x}Sb_x)_2Te_4$**

The band structure evolution upon applied magnetic field and the emergence of Weyl nodes in $Mn(Bi_{1-x}Sb_x)_2Te_4$ can be described by a symmetry-based k·p model. In the nonmagnetic state, the low-energy effective Hamiltonian takes the form[17] in the basis of $|P1_z^+, \uparrow\rangle$, $|P2_z^-, \uparrow\rangle$, $|P1_z^+, \downarrow\rangle$, $|P2_z^-, \downarrow\rangle$, where the $\pm$ and $\uparrow\downarrow$ denote the parity and spin of the corresponding states, respectively:

$$H(\mathbf{k}) = \varepsilon_0(\mathbf{k}) + M_\gamma(\mathbf{k})\tau_z\sigma_0 + A_1 k_z\tau_x\sigma_z + A_2 k_x\tau_x\sigma_x + A_2 k_y\tau_x\sigma_y \quad (1)$$

where $\varepsilon_0(\mathbf{k}) = C + D_1 k_z^2 + D_2(k_x^2 + k_y^2)$, and $M_\gamma(\mathbf{k}) = M_0^\gamma + B_1^\gamma k_z^2 + B_2^\gamma(k_x^2 + k_y^2)$. $\sigma_i$ and $\tau_i$ ($i = x, y, z$) represent the Pauli matrix of the spin and orbit, respectively. $\sigma_0$ and $\tau_0$ stand for the unit matrix. For visualization purposes, the out-of-plane spin-orbit-coupling (SOC) strength $A_1$ is temporarily set to zero. In the nonmagnetic state (Fig. 1, panel i), the system behaves as a topological insulator, where both conduction bands and valence bands are degenerate. The nearly isotropic in-plane dispersion of $Mn(Bi_{1-x}Sb_x)_2Te_4$ allows us to analyze two principal momentum directions, the out-of-plane axis ($k_z$, middle panel) and the in-plane axis ($k_{x,y}$, bottom panel).

Applying a strong magnetic field along z-direction drives the system into FM-z state, where all spin aligns along the field. The description of such state is obtained by adding a perturbative term[18],

$$\Delta H_M(\mathbf{k}) = M_\alpha(\mathbf{k})\tau_0\sigma_z + M_\beta(\mathbf{k})\tau_z\sigma_z + A_3 k_z\tau_x\sigma_0 - A_4 k_x\tau_y\sigma_y + A_4 k_y\tau_y\sigma_x \quad (2)$$

where $M_i(\mathbf{k}) = M_0^i + B_1^i k_z^2 + B_2^i(k_x^2 + k_y^2)$ with $i = \alpha, \beta$. Here, $M_\alpha$ represents the exchange interaction with the orbital magnetic moment of Mn atoms, while $M_\beta$ characterizes bonding-state-dependent coupling strengths. The total Hamiltonian is then given by $H_{FM}(\mathbf{k}) = H(\mathbf{k}) + \Delta H_M(\mathbf{k})$.

Starting from the influence of perturbative terms $M_\alpha$ and $M_\beta$ (panel ii), the external magnetic field lifts the spin degeneracy and induces Zeeman-like splitting through the exchange field, generating a pair of Weyl nodes at $k_{\mathrm{z}} = \pm k_{\mathrm{w}}$. The presence of $M_{\beta}$, dictated by the $\mathcal{T}$-breaking, produces unequal splitting between $|P1_{\mathrm{z}}^{+}\rangle$ and $|P2_{\mathrm{z}}^{-}\rangle$. At $k_{\mathrm{z}} = \pm k_{\mathrm{w}}$, the in-plane dispersion exhibits linear crossings driven by the SOC term $A_2$, accompanied by parabolic bands above and below crossing node. The emergent FM-z phase also introduces an additional in-plane SOC modification term $\pm A_4$ due to the $\mathcal{T}$-breaking (panel iii). The opposite signs of $A_4$ in $\Delta H_{\mathrm{M}}(\mathbf{k})$ generate asymmetric SOC strengths between $|P1_{\mathrm{z}}^{+}, \uparrow\rangle \leftrightarrow |P2_{\mathrm{z}}^{-}, \downarrow\rangle$ and $|P2_{\mathrm{z}}^{-}, \uparrow\rangle \leftrightarrow |P1_{\mathrm{z}}^{+}, \downarrow\rangle$ states.

Stronger coupling in the former indicates a linear in-plane dispersion with higher Fermi velocity, whereas weaker coupling in the latter suppresses dispersion, yielding a nearly flat band (panel iii, bottom). Reintroducing the out-of-plane SOC term $A_1$ couples same-spin, preserving a single pair of Weyl nodes along $k_z$. The resulting type-II Weyl fermions exhibit pronounced in-plane particle-hole symmetry breaking, manifested as a strong velocity asymmetry between conduction and valence bands (panel iv). While mirror, parity, and time-reversal symmetries are extensively studied in topological systems, particle-hole symmetry ($\mathcal{C}$) remains less explored. Here we show that its breaking reshapes Landau-level wavefunctions and underlies the observed circular dichroism.

**Landau Quantization of Type-II Weyl Fermions**

Carrier mobility is crucial for resolving Landau quantization in $MnBi_2Te_4$. To access that, we optimized the growth protocol and introduced Bi→Sb substitution (detailed in Methods). Fig. 2a summarizes the magnetic properties of the as-grown $Mn(Bi_{1-x}Sb_x)_2Te_4$. The ground state is AFM-z, where Mn spins align antiparallel along the out-of-plane direction. When the out-of-plane magnetic field exceeds the spin-flop field $B_{\mathrm{c}}$, the system enters a canted-AFM (cAFM) phase, producing a progressive increase in magnetization. At fields above the saturation field $B_{\mathrm{s}}$, all spins align along the field, driving the system into the FM-z phase with saturated magnetization. The overall magnetic phase diagram undergoes no qualitative changes by increasing the Sb content. Fig. 2b shows the carrier mobility ($\mu$) and carrier density ($n$) extracted from Hall measurements in the FM-z phase. At x ≈ 0.21, the system exhibits a $n-p$ transition and a mobility peak reaching around 2500 $cm^2/Vs$ compared to 100 $cm^2/Vs$ for the intrinsic $MnBi_2Te_4$. The high mobility window rises from the diverging Berry curvature and reduced dynamic mass of Weyl fermion at charge neutrality.

To probe the Landau-level structure, we perform magneto-infrared spectroscopy on $Mn(Bi_{1-x}Sb_x)_2Te_4$. Relative transmission spectra ($T_B/T_R$) are measured in the x-y plane with the magnetic field along the z-axis (Faraday geometry, Fig. 2c), where $T_B$ and $T_R$ denote the transmittance at field $B$ and at reference field (0 T), respectively. Near the spin-flop field $B_c$, a pronounced peak-dip structure (grey arrows) emerges and shifts with increasing field, eventually stabilizing around the saturation field $B_s$. At higher fields, a series of sharper features corresponding to inter-Landau-level transitions ($T_i$) become well resolved. Notably, clear Landau quantization appears only within a narrow Sb range ($0.19 < x < 0.22$), coinciding with the mobility peak. Among these, $x \approx 0.21$ exhibits the sharpest quantization and highest mobility. We therefore focus on this composition (hereafter named as MBST), with other compositions shown in Supplementary information.

To better visualize the phase transition, we convert the spectra to the relative magneto-absorbance, $A_B = -\log_{10}(T_B/T_R)$, as exhibited in Fig. 2d. Below $B_c$, the system remains in the AFM ground state, where the magnetization and exchange interactions keep nearly field-independent. Once the system enters the cAFM phase above $B_c$, the exchange interaction splits and shifts the gap size approximately in proportion to the magnetization[19–21], resulting in spectral weight redistribution near the zero-field gap energy[22] ~180 meV (arrow in Fig. 2c and deep blue in Fig. 2d, more details in Fig. S18), where absorbance is suppressed in one spectral region and enhanced in the others.

To resolve the quantized features more clearly, we analyze the first derivative of the transmission spectra with respect to photon energy, using $B = 3$ T as the reference field (Fig. 2e). As the spins canting towards the z-axis, Landau-level transitions begin to emerge, labeled as $T_1 - T_5$. The Hall mobility predicts the threshold field of about 4 T, in agreement with observation in magneto-infrared spectra. The $\sqrt{B}-$ and $\sqrt{N}-$ dependence of the resonance energy indicate the presence of relativistic fermion (see Extended Data Fig. 2-3).

It is essential to clarify the momentum positions of the optical features, since type-II Weyl fermions differ qualitatively from commonly studied type-I counterparts in their density of states (DOS) distribution[23]. When a magnetic field is applied along the z-axis, the energy spectrum quantizes into a series of one-dimensional Landau bands dispersing with $k_z$. The numerically calculated conduction- and valence-band Landau levels based on k·p model exhibit pronounced differences in both dispersion and level spacing, providing a direct manifestation of particle-hole symmetry breaking in MBST Weyl fermions (Fig. 3a).

Interband Landau-level transitions require both a divergent density of states (DOS) and finite transition matrix elements. At $k_z = 0$, symmetry ensures a divergent DOS. Near the Weyl nodes $\pm k_{\mathrm{w}}$, however, the Landau bands do not form Van Hove singularities, unlike in type-I Weyl systems. However, the slopes of the conduction and valence Landau bands (except $LL_0$) converge (see Extended Data Fig. 4), leading to a predicted[23] band-nesting effect that generates a divergent joint density of states (JDOS). This enables strong inter-band transitions[23,24] at Weyl point. Consequently, both zero-momentum states and states near the Weyl nodes contribute to the observed optical response.

Another necessary condition for observable optical transitions is the presence of non-zero transition matrix elements. The calculated electric-dipole transition matrix is shown in Fig. 3b-c. Optical selection rule reads $\Delta \mathrm{N} = 1$ near the Weyl node. In contrast, only $LL_0 \rightarrow LL_{+1}$ transitions are active for the zero momentum. Taken together, the observable transitions, which require both finite matrix elements and JDOS divergence, are $\mathrm{T}_1 - \mathrm{T}_5$ from $k_{\mathrm{z}} = \pm k_{\mathrm{w}}$ and $\mathrm{T}_0^*$ from $k_{\mathrm{z}} = 0$.

We further employ the linear response formula to reproduce the spectral features of the quantized state (details in Methods). As shown in panel (i) of Fig. 3d, the calculated magneto-optical absorbance, given by the real part of optical conductivity $\sigma_{1,+}$, is symmetric about $k_{\mathrm{z}} = 0$. Panel (ii) presents the $k_{\mathrm{z}}$-summed optical conductivity, demonstrating that most optical features ($\mathrm{T}_1 - \mathrm{T}_5$) originate from the divergent JDOS near the Weyl nodes. In contrast, for the $LL_0 \rightarrow LL_{+1}$ transition, the JDOS diverges at $k_{\mathrm{z}} = 0$ but not at $k_{\mathrm{z}} = \pm k_{\mathrm{w}}$. Therefore, the lowest-energy inter-band Landau-level transition originates from zero momentum and is labeled as $\mathrm{T}_0^*$.

This calculation further enables a quantitative comparison between the theory and experiments. The experimental transition energies extracted directly from the transmission minima in the magneto-infrared spectra, generally agree well with the theoretical model predictions (Fig. 3e). Slight deviation at low field rise can be reproduced by further considering the field-dependent magnetization (Fig. S3 and Extended Data Fig. 5). Our theoretical simulation, based on the k·p model as shown in Fig. 3f, reproduces the overall magneto-infrared response of MBST.

**Circular Dichroisms Generated by Particle-Hole Symmetry Breaking**

To test potential circular dichroism (CD), we developed helicity-resolved magneto-infrared spectroscopy (see Fig. 4b, Fig. S13, Extended Data Fig. 6 and Methods). After reproducing the unpolarized magneto-infrared spectra of MBST (Fig. 4a), which display field-symmetric responses, we introduced a wave plate and wire-grid polarizer to control the incident polarization. The wave plate is optimized for a central wavelength $\lambda_c = 11$ μm, where the transmitted light is nearly circular. By adjusting the relative orientation of the optical elements, we generate $\sigma_+$ and $\sigma_-$ configurations (Fig. 4b). Additional analyzer measurements examined the wavelength-dependent phase retardance (Fig. 4c and supplementary Fig. S12), confirming that although the light is not perfectly circular, it remains strongly dominated by a single helicity over the relevant spectral range. Therefore, the measured CD may only be underestimated, while the sign of the CD is not affected.

Fig. 4d presents the absorption spectra for opposite helicity configurations as $B > 0$. Inter-Landau-level transitions remain well resolved despite the reduced signal-to-noise ratio caused by optical losses from the polarizer and wave plate. The most striking difference compared with Fig. 4a is the spectral asymmetry between the two helicity configurations, which provides direct evidence of CD. $\sigma_+$ photons are strongly absorbed, giving rise to pronounced spectral changes as Landau levels cross the central wavelength, whereas $\sigma_-$ photons exhibit minimal response. This contrast becomes even more apparent in the line plot of transmission spectra (Fig. 4j). When the $T_1$ transition crosses $\lambda_c$ around $B \approx 5.8$-$7.4$ T, where $\sigma_+$ photons produce pronounced transmission minima while $\sigma_-$ photons show nearly no resonance. Away from $\lambda_c$, the incident beam is no longer purely circularly polarized, and a residual admixture of the opposite helicity component appears in the spectra.

A series of additional narrow spectral features are visible near 80 meV, originating from boron impurities in the silicon substrate of the polarizer. The inter-Landau-level transitions of these impurity states scale approximately linearly with magnetic field and exhibit a finite intercept near 83 meV. Importantly, unlike MBST signal, the impurity-related transitions remain nearly symmetric under the two helicity configurations, as shown in Fig. 4h.

The circular dichroism is defined as $\mathrm{CD} \equiv \Delta A = A_{B,\sigma_+} - A_{B,\sigma_-}$, where $A_{B,\sigma_+}$ and $A_{B,\sigma_-}$ are the magneto-absorbances for $\sigma_\pm$ photons[5,7,25,26]. For $B > 0$, pronounced dichroism is observed (right half of Fig. 4e and Fig. 4l), reaching 3240 mdeg (~ 130 mdeg/nm, see Extended Data Fig. 7) around 7-9 μm. The CD exceeds 1000 mdeg covering 6-13 μm, with an average value of 2371 mdeg (~90 mdeg/nm).

The uncertainty of CD is 10-40 mdeg (average 14 mdeg, more detail in Supplementary

Fig. S16). The dichroism increases monotonically with magnetic field and does not peak at the $\lambda_c$, due to the enhancement of oscillator strength of Landau-level transition. The dip near 8.3 μm arises from absorption in the substrate (scotch tape). Further substitution-dependent comparative experiments ($\mathrm{CD} - x$ in Fig. 4n) reveal the presence of such strong CD only around charge-neutral condition (x~0.21). It mimics the mobility peak ($\mu - x$, Fig. 2b) and contrasts the flat magnetization ($M - x$, Fig. 2a). It rules out the conventional magneto-optical effects where CD should follow the magnetization rather than the mobility.

Reversing the magnetic field from $+B$ to $-B$ transforms the system into its time-reversal counterpart, completely inverting the helicity preference. Transitions that previously coupled to $\sigma_+$ photons now couple exclusively to $\sigma_-$ photons, as systematically confirmed in Figs. 4e (left half), 4f, 4g, 4i, 4k, and 4m. Across all four experimental combinations (two helicity configurations × two field orientations), MBST exhibits a single-helicity-dominated response and above-degree broadband CD in the infrared.

In most Landau-quantized materials, particle-hole symmetry breaking only introduces subtle spectral effects[27]. Typically, it causes small splitting between paired inter-band Landau-level transitions such as $\mathrm{LL}_{-1} \rightarrow \mathrm{LL}_{+2}$ and $\mathrm{LL}_{-2} \rightarrow \mathrm{LL}_{+1}$. These splitting lead to weak CD, with its sign often reversing across different transition energies[28], and the overall magnitude of CD remains negligible. Strong and broadband CD is therefore rarely observed in the inter-Landau-level spectroscopy. The experimentally detected strong CD motivated us to calculate the optical conductivity under opposite helicities within the k·p model framework. Fig. 5a shows the real part of calculated optical conductivity $\sigma_{1,\pm}$ corresponding to the absorption of $\sigma_\pm$ polarization. Remarkably, Landau-quantized MBST couples almost exclusively to $\sigma_+$ photon, manifested as the real part of the conductivity, which differs by two to three orders of magnitude between $\sigma_+$ and $\sigma_-$ across the entire spectral range. This extreme helicity-selectivity explains why no particle-hole-induced doublet structures, which would typically accompany such asymmetry, are visible in our magneto-infrared spectra. The CD here is intrinsically broadband due to the suppression of one helicity channel throughout the infrared spectral range, in contrast to tilting- and Fermi-energy-induced circular responses in Weyl semimetal, where CD are typically finite in magnitude, spectrally narrow[29–32], and reaches minimum as the Fermi level approaches the Weyl node (Extended Data Fig. 8).

To pinpoint the origin of strong CD, we expand the ferromagnetic-state Hamiltonian

$H_{\mathrm{FM}}$ (detailed in Methods) near the Weyl node, retaining terms up to first order in the momentum deviation. To quantify the degree of particle-hole asymmetry, a dimensionless parameter $D \equiv 1-(A_2-A_4)/(A_2+A_4)$ is introduced. Fig. 5b shows the evolution of the Weyl cone and the calculated CD of inter-band Landau-level transition as $D$ varies, while keeping all other parameters fixed. In the symmetric case ($D=0$), the optical responses for $\sigma_+$ and $\sigma_-$ photons remain identical. As $D$ increases, particle-hole asymmetry becomes pronounced, and the absorbance difference between the two polarizations grows rapidly. For the experimental value $D\sim 0.93$, MBST exhibits giant and broadband CD in our theory, with $\sigma_+$ absorbance dominating across the entire spectral window.

By establishing the particle-hole symmetry breaking as the underlying mechanism of strong CD, we further analyze its microscopic origin by examining the Landau-level eigenstates in the harmonic-oscillator basis $|\mathrm{n}_{\mathrm{o}}^{\mathrm{s}}\rangle$, where $\mathrm{o}$ and $\mathrm{s}$ denote the orbital and spin indices, respectively. Fig. 5c summarizes the calculated transition matrix between these basis states. The selection rule is given by $\Delta\mathrm{n}=0$, $\Delta\mathrm{o}=\pm 2$, and $\Delta\mathrm{s}=\pm 2$ (for $\sigma_\pm$ absorption), which should not be confused with the $\Delta|\mathrm{M}|=\pm 1$ selection rule for Landau levels. Another type of channel ($\Delta\mathrm{n}=0$, $\Delta\mathrm{o}=\pm 2$, $\Delta\mathrm{s}=\mp 2$) exists, but its matrix elements are two orders of magnitude smaller and thus do not contribute to the optical transitions.

The transition probability between Landau levels ($\mathrm{LL}_{-\mathrm{M}} \rightarrow \mathrm{LL}_{+\mathrm{N}}$) can be written as

$$\left|\langle\psi_{\mathrm{LL}_{+\mathrm{N}}}|\hat{v}_\pm|\psi_{\mathrm{LL}_{-\mathrm{M}}}\rangle\right|^2=\left|\sum_{\mathrm{n,m}} c_{\mathrm{n}}^* c_{\mathrm{m}}\langle\mathrm{n}|\,\hat{v}_\pm|\mathrm{m}\rangle\right|^2 \tag{3}$$

where $c_{\mathrm{n}}$, $c_{\mathrm{m}}$ denote the oscillator-basis projections of the final $\mathrm{N}^{\mathrm{th}}$ Landau band and initial $-\mathrm{M}^{\mathrm{th}}$ Landau band, respectively. Since the magnitudes of the transition matrix elements $\langle\mathrm{n}|\hat{v}_\pm|\mathrm{m}\rangle$ for $\sigma_+$ and $\sigma_-$ channels are identical (Fig. 5c), the pronounced helicity-selectivity must originate from the distribution of the Landau-level eigenstates $c_{\mathrm{n}}^* c_{\mathrm{m}}$ in the harmonic-oscillator basis.

As in Fig. 5d, for $D=0$, Landau levels from conduction and valence bands share identical wavefunction distributions, yielding symmetric transitions and vanishing CD. When particle-hole symmetry is broken ($D\neq 0$), these distributions become asymmetric. In the experimental regime ($D\sim 0.93$), this asymmetry becomes strong: spin-up, parity-even valence components $|\mathrm{n}_+^{\uparrow}\rangle$ in the valence Landau band are nearly depleted, suppressing all $\sigma_-$ transitions. In contrast, the $\sigma_+$ channel remains active, with finite valence components $|\mathrm{n}_-^{\downarrow}\rangle$ and corresponding conduction components $|\mathrm{n}_+^{\uparrow}\rangle$, resulting in strong CD.

Taking $T_1$ as an example, at $D \sim 0.93$ (Fig. 5e), we focus on representative optical transitions $LL_{-1} \rightarrow LL_{+2}$ ($T_{1,+}$, locked to $\sigma_+$) and $LL_{-2} \rightarrow LL_{+1}$ ($T_{1,-}$, locked to $\sigma_-$). Those components responsible for photon absorption are labeled by asterisk and bold box. The $LL_{-2}$ wavefunction has a projection of only 0.00045% onto the $|1_+^{\uparrow}\rangle$ basis. By contrast, the conduction-band levels show no such suppression, with all four basis components present within one order of magnitude. This asymmetry directly impacts the optical response. For $T_{1,-}$, the dominant basis states are $|1_+^{\uparrow}\rangle$ (for initial $LL_{-2}$) and $|1_-^{\downarrow}\rangle$ (for final $LL_{+1}$), as concluded in Fig. 5c and highlighted by bold rectangular borders and asterisks in Fig. 5e. The strong suppression of $|1_+^{\uparrow}\rangle$ in the valence band drives $T_{1,-}$ amplitude effectively to zero, quenching $\sigma_-$ absorption. In contrast, $T_{1,+}$ is mediated by the dominant basis states $|1_-^{\downarrow}\rangle$ (for initial $LL_{-1}$) and $|1_+^{\uparrow}\rangle$ (for final $LL_{+2}$), both of which retain substantial weight in their respective Landau levels, yielding strong $\sigma_+$ photon absorption.

Theoretical calculations therefore support intrinsic, strong and broadband CD of inter-Landau-level transition that extends over a broad spectral range. These results, together with experiments, consistently demonstrate that particle-hole symmetry breaking in MBST enforces a single-helicity channel, giving rise to giant and broadband CD of Landau-level transitions.

**Symmetry Analysis**

From a symmetry perspective, the particle-hole-symmetry-breaking-induced CD can be generalized independently of the microscopic band structure. As exhibited in Fig. 5f, each Landau level has a particle-hole partner ($LL_N \leftrightarrow LL_{-N}$). Likewise, each optical transition has a particle-hole counterpart, e.g., $LL_{-N} \rightarrow LL_{N+1}$ corresponds to $LL_{-N-1} \rightarrow LL_N$. These two transitions are locked to opposite circular polarizations due to angular-momentum conservation. If particle-hole symmetry exists, the two transitions would be identical, and therefore zero CD even in the absence of time-reversal symmetry. When particle-hole symmetry is broken, these paired transitions become inequivalent in energy and intensity, resulting in different absorption of $\sigma_+$ and $\sigma_-$ photon, as shown by Fig. 6. Our analysis can be extended to classical limit without Landau quantization (See Methods and Extended Data Fig. 9).

**Conclusion**

In conclusion, we have uncovered that field-driven-ferromagnetic $Mn(Bi_{1-x}Sb_x)_2Te_4$ hosts strongly particle-hole asymmetric Weyl fermions, with Fermi velocities differing by up to a factor of 16 between conduction and valence bands, that give rise to giant circular dichroism (CD) across a broad spectral range. By developing an optimized growth protocol with controlled Sb substitution, clear Landau quantization are enabled under moderate magnetic fields. Magneto-infrared spectroscopy reveals inter-band Landau-level transitions that are enhanced by the band nesting effect inherent to type-II Weyl dispersions, resulting in a divergent JDOS near the Weyl nodes. Helicity-resolved magneto-infrared spectroscopy, conducted across opposite helicity configurations and magnetic-field directions, consistently proves that $Mn(Bi_{1-x}Sb_x)_2Te_4$ couples exclusively to a single-helicity photon, while the opposite channel is suppressed. The CD exceeds 1000 mdeg throughout 6-13 μm and reaches a maximum at 3240 mdeg (~ 130 mdeg/nm) around 7-9 μm. Our k·p model reproduces the band evolution, Landau-level spectrum, and helicity-selective magneto-optical response, elucidating the microscopic origin of this effect. The simultaneous breaking of particle-hole symmetry and time-reversal symmetry suppresses spin-up, parity-even harmonic-oscillator components in the valence band, which accounts for the extinction of one helicity channel, while leaving the opposite-helicity transitions robust. Numerical calculations show that this mechanism persists in the classical limit. These findings establish particle-hole symmetry breaking as a key and previously overlooked route for engineering chiral optical responses in quantum materials.

**Methods**

**Crystal Growth**

High-quality $Mn(Bi_{1-x}Sb_x)_2Te_4$ single crystals were synthesized using optimized chemical vapor transport (CVT), which helps to suppresses antisite defects and ensures well-ordered Mn sublattices. Precursors were prepared by pre-mixing bismuth grains and antimony lumps with the desired substitution content x at Bi: Sb = (1-x): x, then combining with crushed Mn flakes and Te lumps at $Bi_{1-x}Sb_x$: Mn: Te = 2: y: (3+y), where $2 < y < 3$. The Mn-rich growth environment was employed to compensate for the relatively low transport efficiency of Mn and ensure correct Mn stoichiometry in the final crystals. In addition, a Te-rich atmosphere was introduced to minimize the formation of $Bi_{Mn}$ and $Bi_{Te}$ antisite defects during growth[43]. The entire loading procedure was carried out in an argon atmosphere. The mixture was sealed in a quartz ampule under vacuum, pre-reacted at 900 K for 72 h, finely mixed with $I_2$ as transport agent (~2.5 mg/mL ampule volume), and resealed. The transport procedure was performed in a dual-zone tube furnace with a small temperature difference between the source and sink zones. Notably, a slightly higher temperature and temperature gradient based on Sb content were applied to adapt to the higher crystallization temperatures for Sb-rich samples. As for x around 0.21, a temperature gradient of 30 K (the temperatures of the source and sink zones were set as 865 K and 835 K) was employed to yield shiny, plate-like single crystals in the sink zone after a 14-day reaction.

**Sample Characterization**

The elemental composition of the $Mn(Bi_{1-x}Sb_x)_2Te_4$ single crystals was investigated using energy-dispersive X-ray spectroscopy. Measurements were performed with an electron beam energy set to 12 keV. The obtained elemental ratio of Mn: Bi: Sb: Te = 1: (2-2x): 2x: 4 directly confirmed the accurate stoichiometry of the as-grown crystals.

The field-dependent magnetization ($M-H$ curves) of $Mn(Bi_{1-x}Sb_x)_2Te_4$ single crystals was measured using a Physical Property Measurement System (PPMS) equipped with a vibrating sample magnetometer (VSM). Measurements were performed with an applied magnetic field ranging from 0 to 9 T along the z-axis, enabling the determination of magnetic order transition, especially the spin-flop field ($B_c$) and the saturation field ($B_s$) for various compositions of the samples. For temperature-dependent measurement, an external magnetic field of 500 Oe was applied during field-cooled (FC) processes. The Néel temperature of $Mn(Bi_{1-x}Sb_x)_2Te_4$ ($x \approx 0.21$) is 24.8 K.

Although the overall magnetic phase diagram undergoes no qualitative changes,

increasing the Sb content gradually reduces the saturation moment, the spin-flop field $B_c$, and the saturation field $B_s$ (Fig. 2a). These trends likely arise from shortened Mn-Mn bond length, enhanced magnetic fluctuations, and RKKY-type interactions mediated by Sb-substitution[44], as well as partially compensated magnetization induced by increasing magnetic antisite defects[45]. Nevertheless, the magnetism in our slightly Sb-substituted samples is improved compared with earlier reports, evidenced by a higher Néel temperature (Fig. S10) and higher saturation magnetization.

Electrical transport measurements on $Mn(Bi_{1-x}Sb_x)_2Te_4$ were conducted in a closed-loop superconducting magnet. Prior to measurement, the samples were cleaved by tape exfoliation to ensure a clean, shiny surface. The exfoliated flakes were then affixed to the center of a custom-designed printed circuit board sample holder using GE varnish. Platinum wires were bonded between the sample and the contact pads on the holder using silver conductive paste, allowing for simultaneous measurement of both longitudinal and transverse resistances. The assembled device was mounted onto a sample probe and cooled to a base temperature of 1.6 K. Magnetic field sweeps were conducted between −12 T and +12 T. During all measurements, an alternating current (AC) with a frequency of 17.1717 Hz was supplied by a 6221 current source. The resulting longitudinal and transverse voltage signals were simultaneously detected by two lock-in amplifiers. All measurements were carried out under identical conditions for samples with different compositions. We restrict our Hall analysis to the high-field ferromagnetic regime $|B| > 8$ T (linear regime) for determining the carrier density from the slope of the Hall resistivity $\rho_{xy}$ as a function of magnetic field.

Record-high carrier mobility at x ≈ 0.21 ($\mu \approx 2500$ cm$^2$/V·s) demonstrates the high quality of the crystal. Such high mobility is unusual in magnetic samples, where magnetic scattering typically suppresses coherent transport and Landau quantization. In addition to the reduction of defects, the mobility enhancement mainly originates from the lowered carrier density as the Fermi level approaches the Weyl nodes, where diverging Berry curvature and reduced carrier effective mass strongly suppress backscattering and allow only small-angle scattering[46,47]. However, it is effective only within a narrow substitution window. As the Fermi energy moves away from the Weyl nodes, this suppression mechanism is weakened, leading to a sharp reduction of mobility. Therefore, the condition for Landau quantization $\mu B > 1$ can be satisfied within a very narrow window of Sb substitution. The combination of optimized crystal growth and controlled Sb substitution thus enables the Landau quantization and access to the quantum limit of type-II Weyl fermions in the FM-z phase.

$MnBi_2Te_4$ is chosen for the study due to ferromagnetic Weyl state. It is a layered van der Waals material with a rich phase diagram governed by layer number and magnetic order. Monolayers are ferromagnetic trivial insulators[48]; few-layer films can host quantum anomalous Hall phases with robust chiral edge channels[49–57], or realize axion insulators[58–61] with vanishing optical Hall response[62], due to the preservation of $\mathcal{PT}$ symmetry; and under strong fields, thin films can evolve into high-Chern-number insulators[63–67]. The ground state of bulk $MnBi_2Te_4$ is an antiferromagnetic (AFM-z) topological insulator protected by combined $T\tau_{1/2}$ symmetries[68–71], where $\tau_{1/2}$ denotes a half translation connecting adjacent Mn layers with opposite spin orientations. When driven into the FM-z phase by the external magnetic field, the system undergoes a topological phase transition into a type-II Weyl semimetal hosting only a single pair of Weyl nodes. Slight Sb substitution effectively shifts the Fermi energy towards the Weyl nodes without altering the band topology.

**Magneto-Infrared Spectroscopy**

Magneto-infrared spectra were measured in a home-built 12 T magneto-infrared setup[72], which consists of a Fourier-transform infrared (FTIR) spectrometer and a closed-loop superconducting magnet. A globar was used as the broadband infrared source. The collimated infrared beam from the spectrometer was focused and transmitted through a diamond-windowed optical port of the sample probe. The beam was then guided via a brass pipe to illuminate the sample. Magneto-infrared experiments were performed around liquid helium temperature (4.2 K). The $Mn(Bi_{1-x}Sb_x)_2Te_4$ is exfoliated onto the infrared-transparent tape for the transmission spectra measurement. The exfoliated MBST flakes of various thicknesses cover approximately 30% of the tape surface. The transmission measurements were performed on a macroscopic, round MBST/tape sample of 4.5 mm in diameter. An average thickness of ~63 nm was determined from statistical analysis of the atomic force microscopy measurements (Supplementary Fig. S21). A narrow and field-independent feature near 150 meV originates from the tape. The Faraday geometry was employed, in which the magnetic field was applied parallel to the wavevector of infrared beam and the z-axis (crystallographic c-axis). Light transmitted through the sample was collected by the mercury cadmium telluride (MCT) detector, which is located in an external vacuum chamber. Each spectrum acquisition typically took approximately 2 minutes, and transmission spectra were obtained by Fourier transform.

Zoomed-in magneto-infrared spectra around $B_c$ (Supplementary Fig. S18) clearly shows the band-gap splitting, the gap evolution that scales with magnetization, and the field-induced transparency near the AFM gap energy. These features can be understood in terms of the exchange interaction and are semi-quantitatively reproduced by the proposed four-band model (Supplementary Fig. S19). The results are also consistent with previous studies of the Fermi surface and anomalous Hall effect.

**Helicity-Resolved Magneto-Infrared Spectroscopy**

Helicity-resolved spectroscopy were newly developed in home-built magneto-infrared spectroscopy in ECNU. A circular polarizer constructed by a wide-spectrum wire-grid polarizer (12.5 mm × 12.5 mm, Silicon substrate) and a quarter-wave plate (Φ = 25 mm, $CdGa_2S_4$) was integrated into the home-built magneto-infrared system. It converted the unpolarized infrared beam into left-handed (LCP) or right-handed (RCP) circularly polarized light. To achieve optimal measurement results, the polarization axis of the wire-grid polarizer and the optical axis of the quarter-wave plate were precisely set at an angle of 45°. Prior to the experiment, this angle was calibrated using two linear polarizers and one quarter-wave plate in an independent Fourier-transform spectroscopy measurement. Specifically, the two linear polarizers were placed in the beam path, and their relative orientation was adjusted until a minimum signal of the detector was observed, indicating orthogonal polarization directions. The quarter-wave plate was then inserted between the two linear polarizers. By rotating the wave plate, the angles were identified during a full rotation that yielded the maximum detector signal, corresponding to the angle of ±45° relative to the polarization axis of the polarizer. To determine the purity of circular polarization, the analyzer was further rotated, and the recorded angular dependence of the detector signal confirmed an ellipticity $\xi/\eta = 0.798$ (retardance $\delta \approx 102°$) at the central wavelength, where $\xi$ and $\eta$ represent the length of the semi-minor axis and semi-major axis of the polarization ellipse[73]. More details are given in Supplementary Fig. S12. Once calibrated, the wire-grid polarizer and quarter-wave plate were fixed at this angle to form a stable circular polarizer. Switching between LCP and RCP was achieved by rotating the quarter-wave plate by 90°. To compensate for the optical intensity loss introduced by the polarization optics and to enhance the overall signal-to-noise ratio, we designed a compact "sandwich" module (comprising a ZnSe lens, polarizer, quarter-wave plate, sample, and another ZnSe lens), which was directly mounted onto the sample probe. Additionally, a pair of plane mirrors were replaced with parabolic reflectors to enhance signal collection efficiency. Quarter-wave plates with central wavelength of 11 μm were used to distinguish the broadband response of $Mn(Bi_{1-x}Sb_x)_2Te_4$.

Additional comparative experiments are performed with magnetic field applied along the in-plane direction (Voigt geometry, Supplementary Fig. S20 and Extended Data Fig.

6). In this configuration, Landau-level transitions are suppressed, and the circular dichroism vanishes simultaneously, further confirming that the observed circular dichroism is intrinsically linked to particle-hole symmetry.

**Circular Dichroism Calculation**

Circular dichroism (CD) is defined as[3–6,25,26,74] the absorbance ($A$) differences between left- and right-circular polarized light ($A_{\sigma_+}$ and $A_{\sigma_-}$),

$$\mathrm{CD} \equiv \Delta A = A_{\sigma_+} - A_{\sigma_-} \tag{4}$$

According to the Beer-Lambert law, the absorbance is related to the transmittance ($T$) by,

$$A = -\log_{10}(I/I_0) = -\log_{10} T\,, \tag{5}$$

where $I_0$ is the incident light intensity, and $I$ is the intensity of transmitted light. Thus, CD can be rewritten as,

$$\mathrm{CD} \equiv \Delta A = \log_{10} T_{\sigma_-} - \log_{10} T_{\sigma_+} = \log_{10}\left(T_{\sigma_-}/T_{\sigma_+}\right). \tag{6}$$

In our circular dichroism measurements, the normalized transmission is represented by the ratio of the transmission at the experimental field to that at zero field ($T/T_0$). The absence of $\mathcal{T}$ symmetry guarantees that CD vanishes at zero magnetic field. Accordingly, CD can be further expressed as,

$$\mathrm{CD} = \log_{10}\left(T_{\sigma_-}/T_{\sigma_+}\right) = \log_{10}\left(\frac{T_{\sigma_-}}{T_0}\Big/\frac{T_{\sigma_+}}{T_0}\right) = A_{\sigma_+} - A_{\sigma_-}\,. \tag{7}$$

For practical comparison, CD is often converted into millidegree (mdeg) units,

$$\mathrm{CD_{mdeg}} = 32980 \times \left(A_{\sigma_+} - A_{\sigma_-}\right). \tag{8}$$

In transmission configuration, the general magneto-optical response is described by the complex Faraday angle, $\Theta = \theta_\mathrm{F} + i\eta_\mathrm{F}$, whose real part $\theta_\mathrm{F}$ corresponds to Faraday rotation (circular birefringence) and imaginary part $\eta_\mathrm{F}$ corresponds to circular dichroism[75–77]. Since complex Faraday rotation is conventionally reported in degrees, it is customary to express its imaginary component, i.e., the circular dichroism, in the same unit. We note that Faraday rotation and circular dichroism are related through Kramers-Kronig relations[77–79], but cannot be trivially converted into one another and are therefore not directly comparable. Quantized Faraday rotation has been observed in the 2D surface state of topological insulator at zero frequency limit[61] presenting axion electrodynamics, which differs from the circular dichroism of 3D Weyl fermion at mid-infrared regime.

It is important to emphasize that the measured dichroism away from the central wavelength is underestimated. This limitation arises because no broadband circular polarizers or wave plates are currently available for operation under cryogenic high-field conditions. As a result, single-wavelength quarter-wave plates had to be used, producing reduced polarization purity of incident beam outside the design wavelength. Even at the central wavelength, the dichroism value is also underestimated, because the available commercial infrared wave plates exhibit an ellipticity of 0.798.

Here, the circular dichroism discussed is electronic circular dichroism, rather than vibrational circular dichroism in chiral molecules. It reflects electronic chirality[80] induced by the combined breaking of $\mathcal{T}$ and particle-hole symmetry, rather than geometrically defined chirality.

Meanwhile, several experimental aspects could be improved in future work. Beyond the limited wavelength range of the quarter-wave plate and the imperfect circular polarization it generates, infrared beam leakage also leads to an underestimation of the measured dichroism. In our magneto-infrared spectroscopy experiments, MBST samples were prepared by mechanically exfoliating MBST flakes onto tape. Despite efforts to distribute the flakes as uniformly as possible, unavoidable gaps between flakes allowed part of the incident beam to transmit directly through the tape without interacting with the sample. As a result, in both the $\sigma_+$ and $\sigma_-$ configurations, an additional transmission component was introduced, reducing the measured dichroism relative to the intrinsic property of MBST. To estimate this effect, we consider a simplified model consisting of MBST-covered regions (fraction $x$, thickness 63nm) and uncovered regions $(1-x)$. The measured CD therefore follows $CD_{measured} = \log_{10}\frac{1-x+xT_{\sigma-}}{1-x+xT_{\sigma+}}$. Using the average unpolarized transmission estimated from the optical contrast in Fig. S21c-d, $\frac{T_{\sigma-}+T_{\sigma+}}{2} \approx 70\%$, the corrected CD can be obtained as shown in Extended Data Fig. 7. The correction increases as the coverage decreases as expected. For a coverage of ~50%, the corrected thickness-normalized CD corresponding to the peak value of 3240 mdeg is estimated to be ~ 130 mdeg/nm. All thickness-normalized CD values reported in this work are calculated following this procedure. These observations highlight the need for large-scale and high-quality (especially, higher mobility) Sb-substituted MBT thin films, which would ensure well-defined optical transmission and eliminate undesired leakage. Such samples would also make absolute transmission measurements feasible, thereby enabling a more accurate characterization of the gap evolution and magnetization-induced corrections that are currently difficult to access. Moreover, current experimental infrastructures provide only limited access to circularly polarized infrared spectroscopy under high magnetic fields. We therefore call for the development and integration of helicity-resolved infrared capabilities into high-field facilities, which would accelerate research in chiral optics across the mid- and far-infrared regimes.

**k·p model and numerical calculation**

The experimentally observed Landau-level transitions and circular dichroism are confined to photon energies below approximately 220 meV. Optical transitions in this energy range require initial and final states at the same momentum that straddle the Fermi level with an energy separation smaller than this scale. According to previous density-functional-theory calculations, such low-energy interband transitions exist only in the vicinity of the Γ point of the Brillouin zone.

Constrained by the underlying symmetries, the effective Hamiltonian near Γ in the nonmagnetic phase resembles that of a 3D topological insulator and is given by,

$$H(\mathbf{k}) = \varepsilon_0(\mathbf{k}) + \begin{bmatrix} M_\gamma(\mathbf{k}) & A_1 k_z & 0 & A_2 k_- \\ A_1 k_z & -M_\gamma(\boldsymbol{k}) & A_2 k_- & 0 \\ 0 & A_2 k_+ & M_\gamma(\mathbf{k}) & -A_1 k_z \\ A_2 k_+ & 0 & -A_1 k_z & -M_\gamma(\mathbf{k}) \end{bmatrix} \tag{9}$$

where $k_\pm = k_x \pm i k_y$. Once the system enters the ferromagnetic phase under a sufficiently strong magnetic field (along z direction), the Hamiltonian is modified by the additional perturbation $\Delta H_M$, which respects the symmetry breaking induced in the FM phase[18] and reads,

$$\Delta H_M(\mathbf{k}) = \begin{bmatrix} M_\alpha(\mathbf{k}) + M_\beta(\mathbf{k}) & A_3 k_z & 0 & A_4 k_- \\ A_3 k_z & M_\alpha(\mathbf{k}) - M_\beta(\mathbf{k}) & -A_4 k_- & 0 \\ 0 & -A_4 k_+ & -M_\alpha(\mathbf{k}) - M_\beta(\mathbf{k}) & A_3 k_z \\ A_4 k_+ & 0 & A_3 k_z & -M_\alpha(\mathbf{k}) + M_\beta(\mathbf{k}) \end{bmatrix} \tag{10}$$

In addition to $\Delta H_M$, Landau quantization perpendicular to the magnetic field plane is also incorporated. Under the Landau gauge, we follow the Peierls substitution $\boldsymbol{\pi} = \hbar\mathbf{k} + e\mathbf{A}$, and define the ladder operators,

$$a = \frac{l_B}{\sqrt{2}\hbar}\left(\pi_x - i\pi_y\right)$$
$$a^\dagger = \frac{l_B}{\sqrt{2}\hbar}\left(\pi_x + i\pi_y\right) \tag{11}$$

where $l_B = \sqrt{\hbar/eB}$ is the magnetic length. To proceed, we adopt a four-component harmonic oscillator basis and perform numerical diagonalization to calculate the Landau band along $k_z$ under magnetic fields. The Hamiltonian matrix corresponding to the eigenvalue equation is written as,

$$H_{LL} = \begin{bmatrix} H_{00} & H_{10}^\dagger & 0 & 0 \\ H_{10} & H_{11} & H_{21}^\dagger & 0 \\ 0 & H_{21} & H_{22} & \ddots \\ 0 & 0 & \ddots & \ddots \end{bmatrix} \tag{12}$$

with

$$H_{nn} = \begin{bmatrix} W_1 + \frac{2V_1}{l_B^2} n & (A_1 + A_3)k_z & 0 & 0 \\ (A_1 + A_3)k_z & W_2 + \frac{2V_2}{l_B^2} n & 0 & 0 \\ 0 & 0 & W_3 + \frac{2V_3}{l_B^2} n & (-A_1 + A_3)k_z \\ 0 & 0 & (-A_1 + A_3)k_z & W_4 + \frac{2V_4}{l_B^2} n \end{bmatrix},$$

$$H_{n+1,n} = \begin{bmatrix} 0 & 0 & 0 & 0 \\ 0 & 0 & 0 & 0 \\ 0 & (A_2 - A_4)\frac{\sqrt{2(n+1)}}{l_B} & 0 & 0 \\ (A_2 + A_4)\frac{\sqrt{2(n+1)}}{l_B} & 0 & 0 & 0 \end{bmatrix}, \qquad (13)$$

where $n \geq 0$, $W_{\mathrm{ii}} = P_0^{\mathrm{ii}} + P_1^{\mathrm{ii}} k_z^2 + \frac{V_{\mathrm{ii}}}{l_B^2}$ (ii = 1,2,3,4), and $V_{\mathrm{ii}}$ (ii = 1,2,3,4) denotes the linear combination of parameters in Eq. (1)-(2), given as,

$$\begin{aligned} V_1 &= B_2^{\alpha} + B_2^{\beta} + B_2^{\gamma} + D_2 \\ V_2 &= B_2^{\alpha} - B_2^{\beta} - B_2^{\gamma} + D_2 \\ V_3 &= -B_2^{\alpha} - B_2^{\beta} + B_2^{\gamma} + D_2 \\ V_4 &= -B_2^{\alpha} + B_2^{\beta} - B_2^{\gamma} + D_2 \end{aligned} \qquad (14)$$

With a sufficiently large cutoff $N_c = 200$, the calculated Landau bands are well converged. Most of the parameters in the model are directly adopted from density functional theory (DFT) results reported in a previous study[18], with minor adjustments to better match the experiments through a multi-parameter fitting procedure.

To capture the inter-band Landau-level transition features observed in magneto-infrared spectroscopy, we evaluate the optical conductivity using the Kubo formula

$$\sigma_{1,\pm}(\omega, B, k_z) \propto \mathrm{Re}\left[ iB \sum_{\mathrm{M,N}} \int \frac{\mathrm{d}k_z}{2\pi} \left( \frac{f(E_{\mathrm{M}}) - f(E_{\mathrm{N}})}{E_{\mathrm{N}} - E_{\mathrm{M}}} \cdot \frac{\left|\langle \psi_{\mathrm{N}} | \hat{v}_{\pm} | \psi_{\mathrm{M}} \rangle\right|^2}{E_{\mathrm{N}} - E_{\mathrm{M}} - \hbar\omega + i\Gamma} \right) \right] \qquad (15)$$

where $\sigma_{1,\pm}(\omega, B, k_z)$ denotes the real part of optical conductivity, which is directly proportional to the magneto-absorbance. $f(E_{\mathrm{i}}) = 1/\left[e^{(E_{\mathrm{i}} - E_{\mathrm{F}})/k_{\mathrm{B}}T} + 1\right]$, $\mathrm{i} = \mathrm{N}, \mathrm{M}$, is the Fermi-Dirac distribution function, and $\Gamma$ represents the linewidth broadening. The symbols "+" and "–" distinguish the absorption of opposite circularly polarized light and are associated with the velocity operator $\hat{v}_{\pm} = (\hat{v}_x \pm i\hat{v}_y)/\sqrt{2}$ in transition matrix elements. According to Fermi's golden rule, the transition probability (corresponding to the real part of optical conductivity or absorption coefficient) is governed by the electric-dipole transition matrix elements and JDOS (both are $k_z$-dependent). We verify that the chosen momentum truncation does not alter the number or position of peaks in the optical conductivity after integration, as no divergence occurs in the JDOS beyond the considered region. All calculations are performed at zero temperature, with the Fermi energy $E_{\mathrm{F}}$ chosen to agree with experimental results.

In the ferromagnetic phase, the Hamiltonian is further expanded to first order in the momentum deviation $\mathbf{q}$ = ($q_x$, $q_y$, $q_z$) around the Weyl node $\mathbf{K}$ = (0, 0, $k_w$), to highlight the key parameters responsible for particle–hole asymmetry, expressed as

$$H_{\mathrm{FM}}(\mathbf{q}) = H(\mathbf{q}) + \Delta H_{\mathrm{M}}(\mathbf{q})$$
$$= \begin{bmatrix} Q_1(\mathbf{q}) & (A_1 + A_3)(k_w + q_z) & 0 & (A_2 + A_4)(q_x - iq_y) \\ (A_1 + A_3)(k_w + q_z) & Q_2(\mathbf{q}) & (A_2 - A_4)(q_x - iq_y) & 0 \\ 0 & (A_2 - A_4)(q_x + iq_y) & Q_3(\mathbf{q}) & (-A_1 + A_3)(k_w + q_z) \\ (A_2 + A_4)(q_x + iq_y) & 0 & (-A_1 + A_3)(k_w + q_z) & Q_4(\mathbf{q}) \end{bmatrix} \quad (16)$$

where $Q_{\mathrm{ii}}(\mathbf{q}) = P_0^{\mathrm{ii}} + P_1^{\mathrm{ii}} k_w^2 + 2P_1^{\mathrm{ii}} k_w q_z$ with $\mathrm{ii} = 1,2,3,4$, and $P_{0,1}^{\mathrm{ii}}$ denotes the linear combination of parameters in Eq. (1)-(2), which reads

$$\begin{aligned} P_0^1 &= M_0^\alpha + M_0^\beta + M_0^\gamma + C \\ P_1^1 &= B_1^\alpha + B_1^\beta + B_1^\gamma + D_1 \\ P_0^2 &= M_0^\alpha - M_0^\beta - M_0^\gamma + C \\ P_1^2 &= B_1^\alpha - B_1^\beta - B_1^\gamma + D_1 \\ P_0^3 &= -M_0^\alpha - M_0^\beta + M_0^\gamma + C \\ P_1^3 &= -B_1^\alpha - B_1^\beta + B_1^\gamma + D_1 \\ P_0^4 &= -M_0^\alpha + M_0^\beta - M_0^\gamma + C \\ P_1^4 &= -B_1^\alpha + B_1^\beta - B_1^\gamma + D_1 \end{aligned} \quad (17)$$

In this low-energy expansion, the two SOC parameters, $A_2$ (original in-plane SOC) and $A_4$ (FM-induced in-plane SOC term), play competing roles. Their opposite signs split the effective couplings between the $|P1_z^+, \uparrow\rangle \leftrightarrow |P2_z^-, \downarrow\rangle$ and $|P2_z^-, \uparrow\rangle \leftrightarrow |P1_z^+, \downarrow\rangle$ channels. As $A_4$ approaches $A_2$, the coupling for $|P2_z^-, \uparrow\rangle \leftrightarrow |P1_z^+, \downarrow\rangle$ is suppressed toward zero, flattening one branch of the dispersion. Meanwhile, the coupling for $|P1_z^+, \uparrow\rangle \leftrightarrow |P2_z^-, \downarrow\rangle$ is enhanced, generating a sharply linear band crossing with a Fermi velocity roughly proportional to ($A_2$ + $A_4$). This imbalance leads to a pronounced particle-hole asymmetry, reflected in the strongly unequal in-plane Fermi velocities of conduction and valence bands. The degree of particle-hole asymmetry $D$ is therefore given by

$$D \equiv 1 - (A_2 - A_4)/(A_2 + A_4) \quad (18)$$

As intermediate steps in our calculation, the Landau-level spectra and the $k_z$-dependent conductivity are shown in Fig. S6.

Taken together with previous angle-resolved photoemission spectroscopy (ARPES)[81]

and transport reports, these results indicate that Sb substitution $x$ does not induce abrupt changes in the band structure, topology, or magnetic order over the wide range of $x$ relevant to this work, indicating the applicability of the symmetry-based k·p model upon Sb-substitution.

**Symmetry Analysis**

In the above k·p model, the mechanism by which particle-hole symmetry breaking generates circular dichroism can be generalized, on symmetry grounds, to more generic Landau-quantized systems. As exhibited in Fig. 5f, each Landau level has a particle-hole partner, so do the optical transitions. These transitions are locked to opposite circular polarizations due to angular-momentum conservation. If particle-hole symmetry exists, the two transitions would be identical by the definition of particle-hole symmetry, implying an identical optical response to $\sigma_+$ and $\sigma_-$ absorption, and therefore zero circular dichroism even in the absence of time-reversal symmetry. When particle-hole symmetry is broken, these paired transitions become inequivalent in energy and intensity, resulting in non-zero circular dichroism. In MBST, the strong particle-hole asymmetry makes these differences particularly large. This symmetry-based argument is insensitive to microscopic band-structure details. Therefore, in Fig. 5a, difference between the optical conductivity for $\sigma_+$ and $\sigma_-$ channels exists beyond Weyl node momentum. However, a divergence of the joint density of states arises from the Weyl point strongly amplifies the optical response to circularly polarized light. The comparison of the circular dichroism value and thickness normalized value in our work with existing literatures are provided in Extended Data Fig. 1.

The combined time-reversal symmetry ($\mathcal{T}$) breaking and particle-hole symmetry ($\mathcal{C}$) breaking leads to an unusual broadband and strong circular dichroism response in the infrared regime, compared with cases involving only parity symmetry ($\mathcal{P}$) breaking or $\mathcal{T}$ breaking cases (Fig. 6). Even with strong $\mathcal{T}$ breaking induced by high magnetic field, circular dichroism should vanish exactly in the presence of $\mathcal{C}$ symmetry for Dirac and Weyl semimetals, which underscores the key role of $\mathcal{C}$ asymmetry for the pronounced circular dichroism. We expect that strong circular dichroism is likely to be a universal phenomenon across topological condensed matter under Landau quantization, since particle-hole asymmetry naturally exists. In topological materials, band inversion typically involves states with different orbital origins, which inherently leads to particle-hole asymmetry, as exemplified by the Bi- and Te-derived bands in MBST. As shown in Fig. 5b, even a modest degree of particle-hole symmetry breaking can generate substantial circular dichroism, producing orders-of-magnitude differences in absorption for opposite helicities. Thus, our theory, symmetry analysis, and experimental results suggest that a wide class of topological matters may exhibit circular dichroism, which calls for further helicity-resolved measurements across different material platforms.

**Circular Dichroism in Classical-Limit**

Circular dichroism remains pronounced outside the field and spectral window of the Landau-level transitions (e.g., ~200 meV at $B \approx 3$ T), indicating that the strong-field CD is a quantized manifestation of its classical-limit counterpart. To extend the discussed mechanism beyond Landau level, we performed the same model calculation as in Fig. 5b, but without imposing Landau quantization. The numerical results (Extended Data Fig. 8) for classical type-II Weyl band in MBST exhibit vanishing CD in particle-hole symmetric limit and strong and broadband CD emerges for asymmetry case. This evolution closely mirrors the trend obtained in the Landau-quantized analysis, demonstrating that the fundamental mechanism is identical in both cases: particle-hole symmetry breaking in a magnetic Weyl system generates circular dichroism, regardless of Landau quantization. So that Landau quantization is not pre-requisite to access such strong CD. In this sense, the Landau-level picture represents the quantized limit of the same underlying mechanism, whereas the non-quantized calculation corresponds to its classical limit.

Unlike cyclotron resonance in doped systems, which yield strong but spectrally narrow circular dichroism due to the activation of single Landau-level index at a given magnetic field because of Pauli blocking (Extended Data Fig. 9), the circular dichroism reported here arises from multiple interband Landau-level transitions governed by the optical transition matrix elements. As a result, the reported CD here is intrinsically broadband, in contrast to conventional intraband transition.

**Data availability**

The data generated in this study are provided in the Source Data file. All other data that support the findings of this study are available from the corresponding author upon request. Source data are provided with this paper.

**Methods-only references**

## Figures

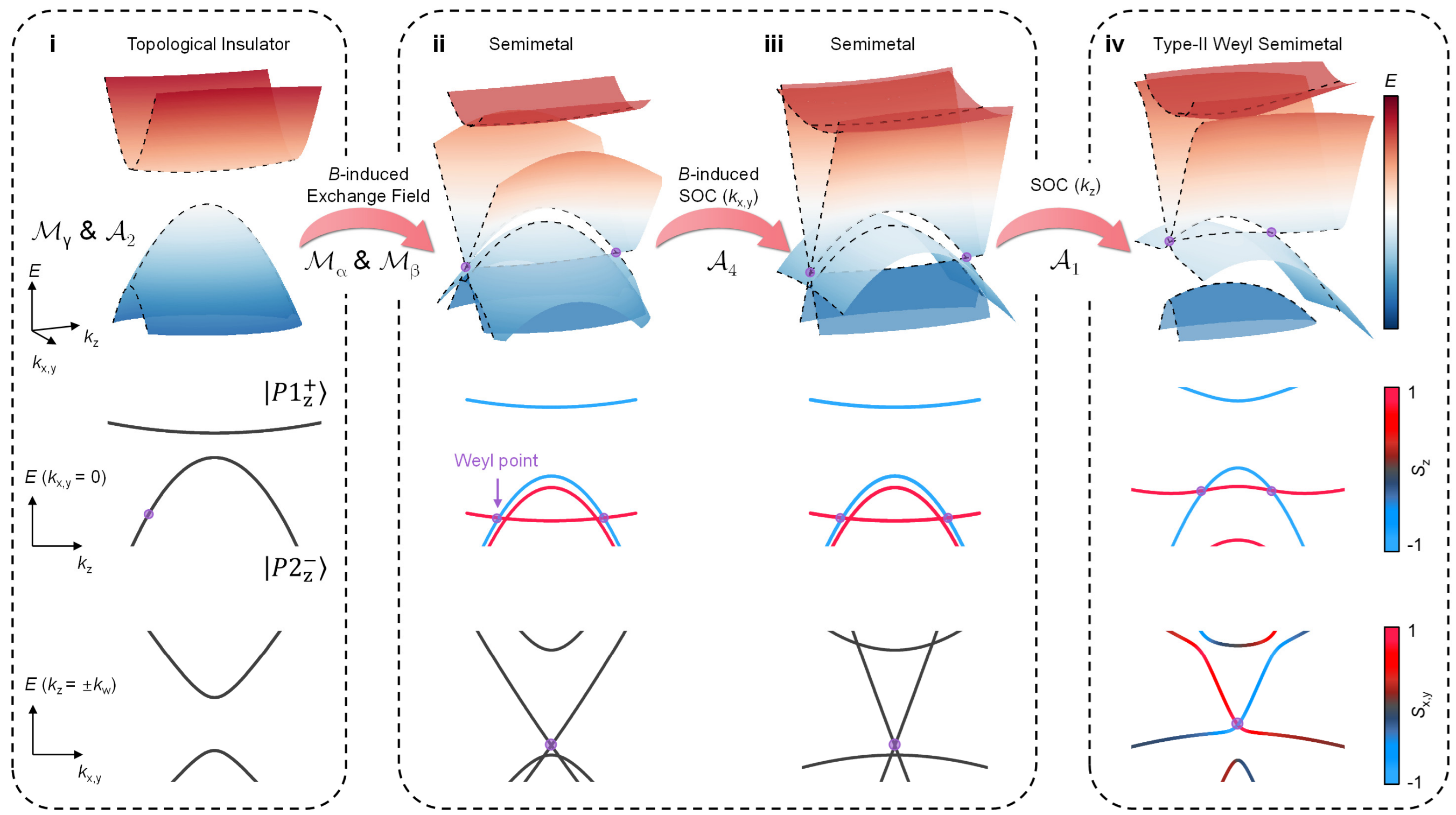

**Fig. 1 | Origin of Weyl nodes and particle-hole symmetry breaking in $Mn(Bi_{1-x}Sb_x)_2Te_4$.** Electronic band structures in different magnetic states, calculated using a symmetry-based k·p model. (i) In the nonmagnetic state, the spin-degenerate states $|P1_z^+\rangle$, $|P2_z^-\rangle$ originate from Bi and Te layers, respectively. (ii) A strong out-of-plane magnetic field polarizes all spins, inducing a ferromagnetic phase. Exchange interactions $M_\alpha$ and $M_\beta$ lift spin degeneracy and generate Weyl nodes along the $k_z$ axis. (iii) The broken $\mathcal{T}$ and $C_{2x}$ symmetries in ferromagnetic phase introduce an additional in-plane SOC term $A_4$. Its opposite contributions to the orbital-spin channels ($|P1_z^+, \uparrow\rangle \leftrightarrow |P2_z^-, \downarrow\rangle$) and ($|P2_z^-, \uparrow\rangle \leftrightarrow |P1_z^+, \downarrow\rangle$) produce asymmetric band dispersion, the former becomes more linear with enhanced velocity, while the latter flattens toward a quasi-flat band. (iv) Reintroducing out-of-plane SOC term $A_1$ opens gaps between the states with the same spin, yielding a single pair of nodes with pronounced in-plane particle-hole asymmetry. The color scale in middle and bottom panels represents the spin expectation ($S_z$ and $S_{x,y}$) along the corresponding direction.

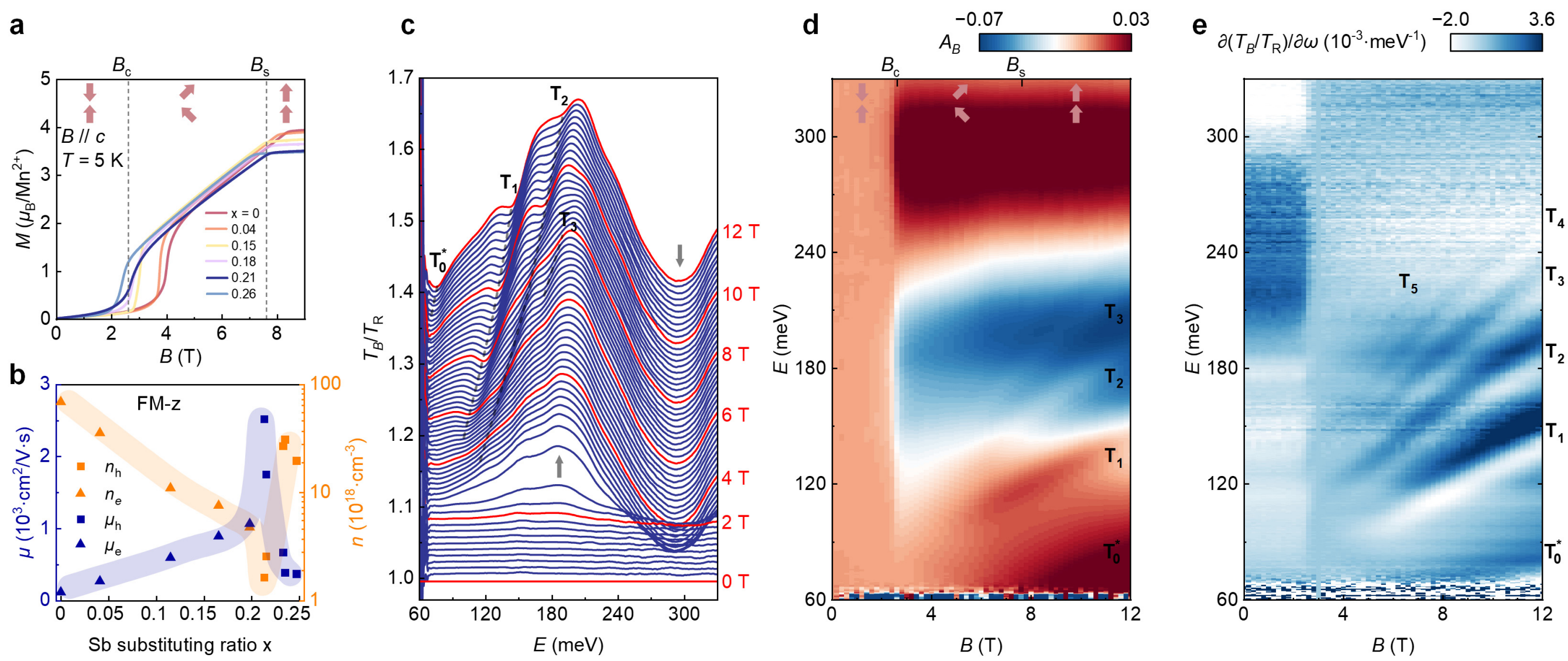

**Fig. 2 | Magneto-infrared spectroscopy of $Mn(Bi_{1-x}Sb_x)_2Te_4$. a,** Magnetization measurements. The critical fields $B_c$ and $B_s$ mark the critical field of transitions from antiferromagnetic (AFM-z) to canted-AFM (cAFM) and then to ferromagnetic (FM-z) phases for x ≈ 0.21. All compositions follow a similar magnetic phase diagram, with systematic evolution across substitution. **b,** Transport measurements at 2 K. The Hall mobility peaks near x ≈ 0.21, exhibiting a 20-fold enhancement over that of intrinsic $MnBi_2Te_4$. This enhancement enables Landau quantization in the ferromagnetic regime. **c,** Magneto-infrared transmission spectra. Three distinct features emerge: a sharp spectral jump near $B_c$, a broad peak-dip feature (grey arrows) that gradually evolves between $B_c$ and $B_s$, but becomes stationary above $B_s$, and a series of narrow, field-dependent transmittance minima at high fields, labeled as $T_0^* - T_3$. **d,** Magneto-absorbance spectra. In the AFM phase ($B < B_c$), the spectra remain nearly unchanged. Above $B_c$, spectral weight redistributes between 200-300 meV with increasing magnetization due to exchange-interaction-driven band shifts. **e,** Differential transmission spectra. The high-field regime reveals a series of inter-band Landau-level transitions. Their nonlinear $\sqrt{B}$ scaling, zero-field intercept, and decreasing mode spacing with energy collectively indicate Landau quantization of massless relativistic fermions.

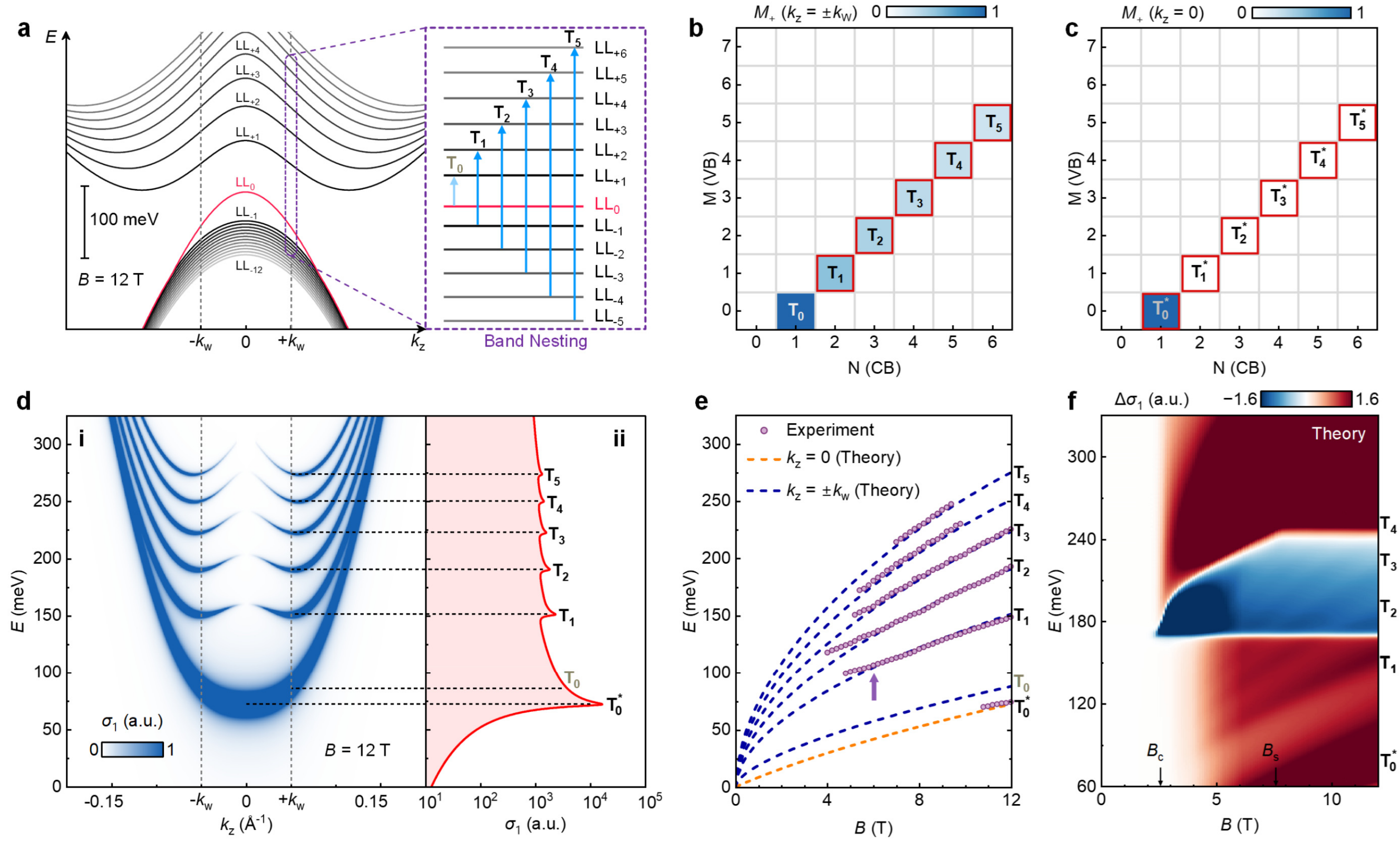

**Fig. 3 | Origin of Landau-level transitions. a,** Landau-level (LL) dispersions calculated from a symmetry-based k·p model. Parameters fitted by magneto-infrared spectroscopy are in agreement with prior DFT results[18]. At $k_z = 0$, the band extrema give rise to Van Hove singularities. Near the Weyl nodes momenta $\pm k_W$, although the DOS does not diverge, band nesting strongly enhances the JDOS, enabling pronounced inter-Landau-level transitions ( $T_1 - T_5$ , blue arrows). **b-c,** Normalized optical transition matrix $|\langle\psi_N|\hat{v}_+|\psi_M\rangle|^2$ for LL-M → LL+N at $k_z = \pm k_W$ (labelled as $T_M$) and $k_z = 0$ ($T_M^*$). The vertical axis represents the M$^{th}$ valence-band Landau level (initial states), and the horizontal axis represents the N$^{th}$ conduction-band Landau level (final states). Transitions associated with divergent JDOS, which may contribute to the experimental spectra, are highlighted with red borders. At $k_z = \pm k_W$, the matrix element allows transitions with $\Delta N = 1$, while at $k_z = 0$, only the LL0 → LL+1 transition (labeled as $T_0^*$) is permitted. **d,** Theoretical optical conductivity. Left panel: real part of conductivity as a function of $k_z$. Right panel: $k_z$-integrated optical conductivity. Black dashed lines connect the JDOS divergence points to their corresponding spectral peaks. The $T_0$ transition does not contribute to optical absorption due to the lack of JDOS divergence. It is also emphasized that the real part of the conductivity $\sigma_{1,\pm}$ should not be confused with $\sigma_\pm$, which describes circularly polarized photons. **e,** Comparison between theory and experiment. The dashed and dotted curves represent theoretical transition energies originating from $k_z = \pm k_W$ and $k_z = 0$, respectively. Transitions $T_1 - T_5$ originate from Weyl nodes, whereas the $T_0^*$ rises from the zero-momentum channel. The low-field deviation denoted by the purple arrow is well explained by including gradual spin canting from the AFM state into the model. **f,** Simulated magneto-infrared spectrum reproducing the key features observed experimentally (Fig. 2d).

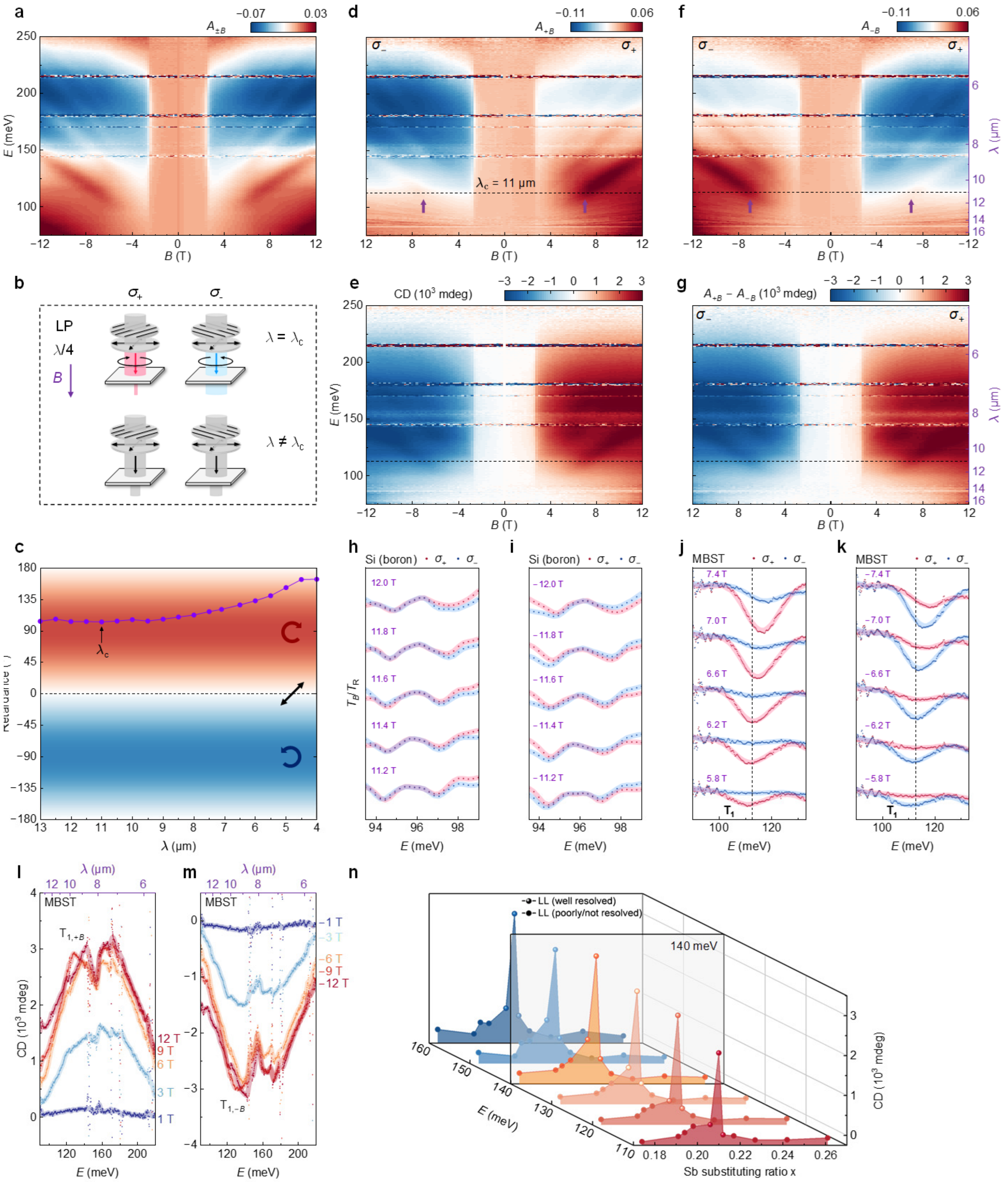

**Fig. 4 | Helicity-resolved magneto-infrared spectroscopy of Mn(Bi$_{1-x}$Sb$_x$)$_2$Te$_4$. a,** Unpolarized magneto-infrared absorption spectra of MBST flakes, showing field-symmetric inter-band Landau-level transitions. **b,** Experimental configuration for helicity-resolved magneto-infrared spectroscopy. Wire-grid polarizer (LP) and quarter-wave plate ($\lambda/4$, designed for $\lambda_c$ = 11 μm) generate left- or right-handed circularly polarized light at the design wavelength. **c,** Phase retardance the incident photon as a function of wavelength,

indicating that the polarization state of incident beam is always dominated by a single helicity. **d,f,** Helicity-resolved magneto-absorbance in positive and negative magnetic field. The strong contrast between the $\sigma_+$ and $\sigma_-$ configuration reveals inter-band Landau-level transitions with pronounced helicity dependence, while additional low-energy absorption near 80 meV originates from Landau quantization of the boron impurity in Si polarizer substrate. **e,** Circular dichroism of MBST at opposite magnetic field, reaching maximum value over $3\times10^3$ mdeg. **g,** The difference in absorption under opposite magnetic fields further supports the presence of above-degree circular dichroism. **h,i,** Helicity-dependent magneto-transmission spectra around boron impurity level of the silicon polarizer substrate. Red and blue dots represent experimental data for $\sigma_+$ and $\sigma_-$, respectively, and light-colored solid lines denote smoothed curves. These impurity absorptions remain nearly helicity-symmetric and are therefore invisible in panel **e**, in sharp contrast to the strong intrinsic dichroism of MBST. **j,k,** Magneto-transmittance near the design wavelength of the wave plate. At the central wavelength, transmission minima appear almost exclusively in one helicity channel, exhibiting pronounced helicity preference, particularly when a Landau-level transition cross $\lambda_c$. A vertical offset has been applied for better comparison. **l,m,** Field-dependent circular dichroism spectra. The response generally exceeds 1000 mdeg within 6-13 μm, and reaches a maximum value of 3240 mdeg (~ 130 mdeg/nm) around 7-9 μm. Switching the direction of the magnetic field exhibits a similar response with reversed sign as expected. **n,** Three-dimensional CD-$x$ stacked plot showing the evolution of CD as a function of substitution ratio at different photon energies. Spherical markers indicate doping levels where Landau levels are well resolved, whereas dot markers denote compositions for which Landau quantization is weak or not resolvable.

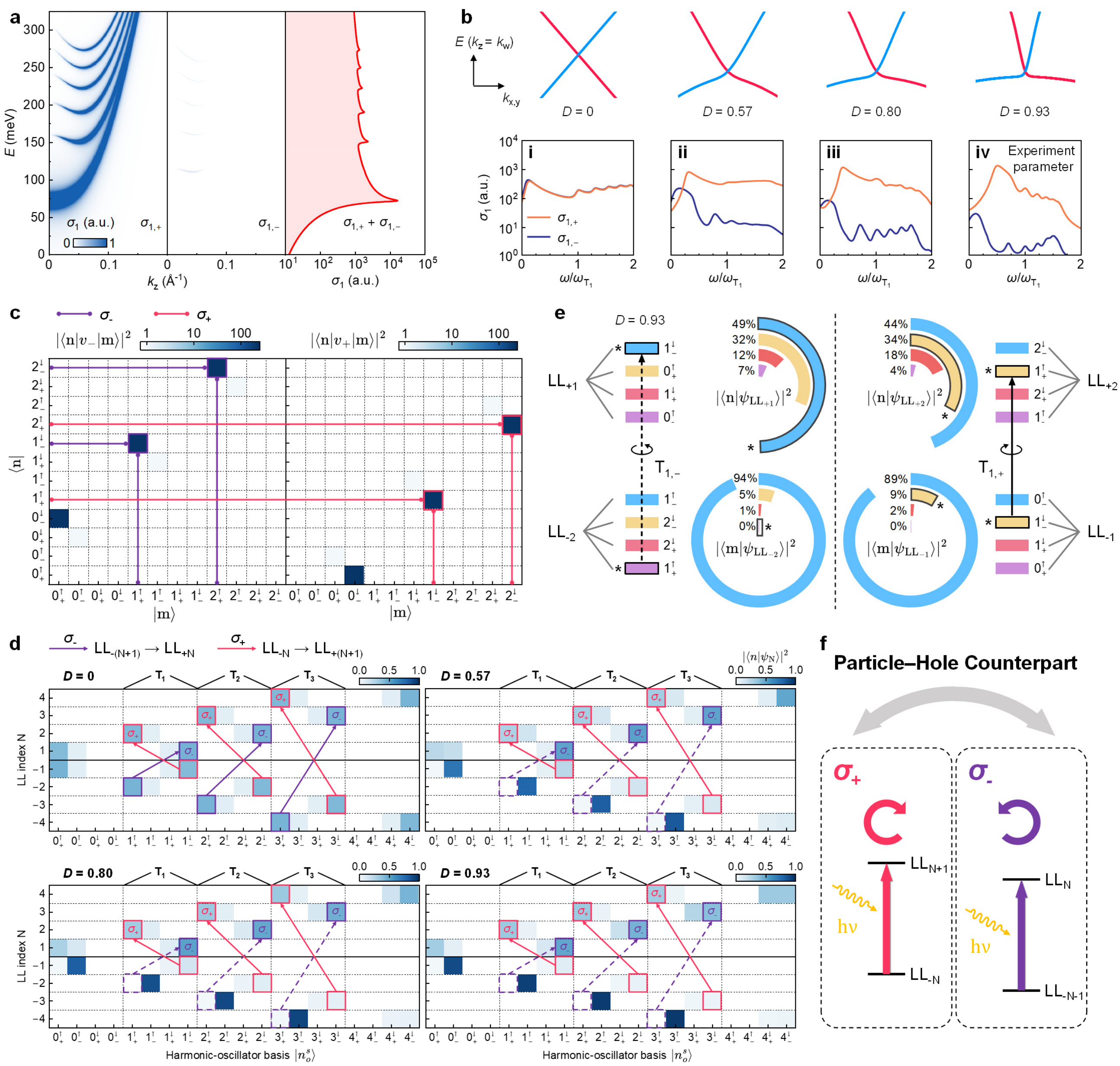

**Fig. 5 | Mechanism of circular dichroism. a,** Real part of magneto-optical conductivity ($\sigma_{1,\pm}$) for opposite helicities. Summed conductivity is dominated by the $\sigma_+$ channel, as absorption for $\sigma_-$ photons is nearly vanishing across the spectral range. **b,** Evolution of the band structure and corresponding circular dichroism of Landau-level transition as the asymmetry parameter ($D$) varies. Band dispersion and optical conductivity are calculated using a low-energy expansion near the Weyl node. The photon energy is normalized by the energy of $T_{1,+}$. As $D$ increases from 0 (symmetric limit) to 0.93 (experimental value), the Weyl cone evolves from particle-hole symmetric to strongly asymmetric. Optical response transitions from identical absorption for both helicities to exclusive coupling with $\sigma_+$ photon, revealing a direct link between particle-hole symmetry breaking and circular dichroism. **c,** Transition matrix $|\langle n|\hat{v}_\pm|m\rangle|^2$ on the harmonic oscillator basis for a single Weyl cone. The eigenstates $|n\rangle$ and $|m\rangle$ denote final and initial states, respectively. Red and purple solid lines indicate transitions activated by $\sigma_+$ and $\sigma_-$ photon. The matrix elements for both helicities are identical. **d,** Evolution of Landau-level wavefunction on the harmonic-

oscillator bases with asymmetry parameter $D$. The arrows indicate all transitions allowed by the selection rules of the harmonic-oscillator basis; the arrow colors represent optical transitions of opposite helicity. Solid arrows denote strong transitions, while dashed arrows denote weak or forbidden transitions. The purple box indicates the dominant basis $\langle n_{+}^{\uparrow}|$ of initial Landau level for $\sigma_{-}$ channel ($LL_{-(N+1)} \rightarrow LL_{+N}$). The red box indicates the dominating basis $\langle n_{-}^{\downarrow}|$ of initial Landau level for $\sigma_{+}$ channel ($LL_{-N} \rightarrow LL_{+N+1}$). At $D = 0$, symmetry enforces identical wavefunction distributions in the conduction and valence bands and identical transitions for opposite helicities. As $D$ increases, the distribution of basis states for conduction-band Landau levels ($N > 0$, upper half) and valence-band Landau levels ($N < 0$, lower half) becomes increasingly different. Those basis responsible for $\sigma_{-}$ chanel, i.e. $\langle 1_{+}^{\uparrow}|$ together with all spin-up, parity-even components in the valence band, becomes nearly extinguished (purple dashed square) in the experimental case ($D \sim 0.93$), whereas the $\sigma_{+}$ channel remains unaffected. Therefore, the redistribution of basis composition with increasing $D$ suppresses all the $\sigma_{-}$ transition (purple dashed arrow) while leaving the $\sigma_{+}$ transition (red solid arrow) essentially unchanged, giving rise to pronounced helicity selectivity across broad energy range. **e,** Example of the $T_1$ transition. Basis $\langle 1_{+}^{\uparrow}|$ is responsible for the $\sigma_{-}$ channel of $T_1$, but $\langle 1_{+}^{\uparrow}|$ nearly vanishes in the valence band, leading to suppression of the $\sigma_{-}$ channel. **f.** Symmetry analysis of particle-hole symmetry breaking-induced circular dichroism. From the symmetry perspective, each Landau level has a particle-hole partner ($LL_{N} \leftrightarrow LL_{-N}$). Likewise, each optical transition has a particle-hole counterpart, e.g., $LL_{-N} \rightarrow LL_{N+1}$ corresponds to $LL_{-N-1} \rightarrow LL_{N}$. These two transitions are locked to opposite circular polarizations due to angular-momentum conservation. Therefore, the particle-hole symmetry enforces zero circular dichroism, and vice versa.

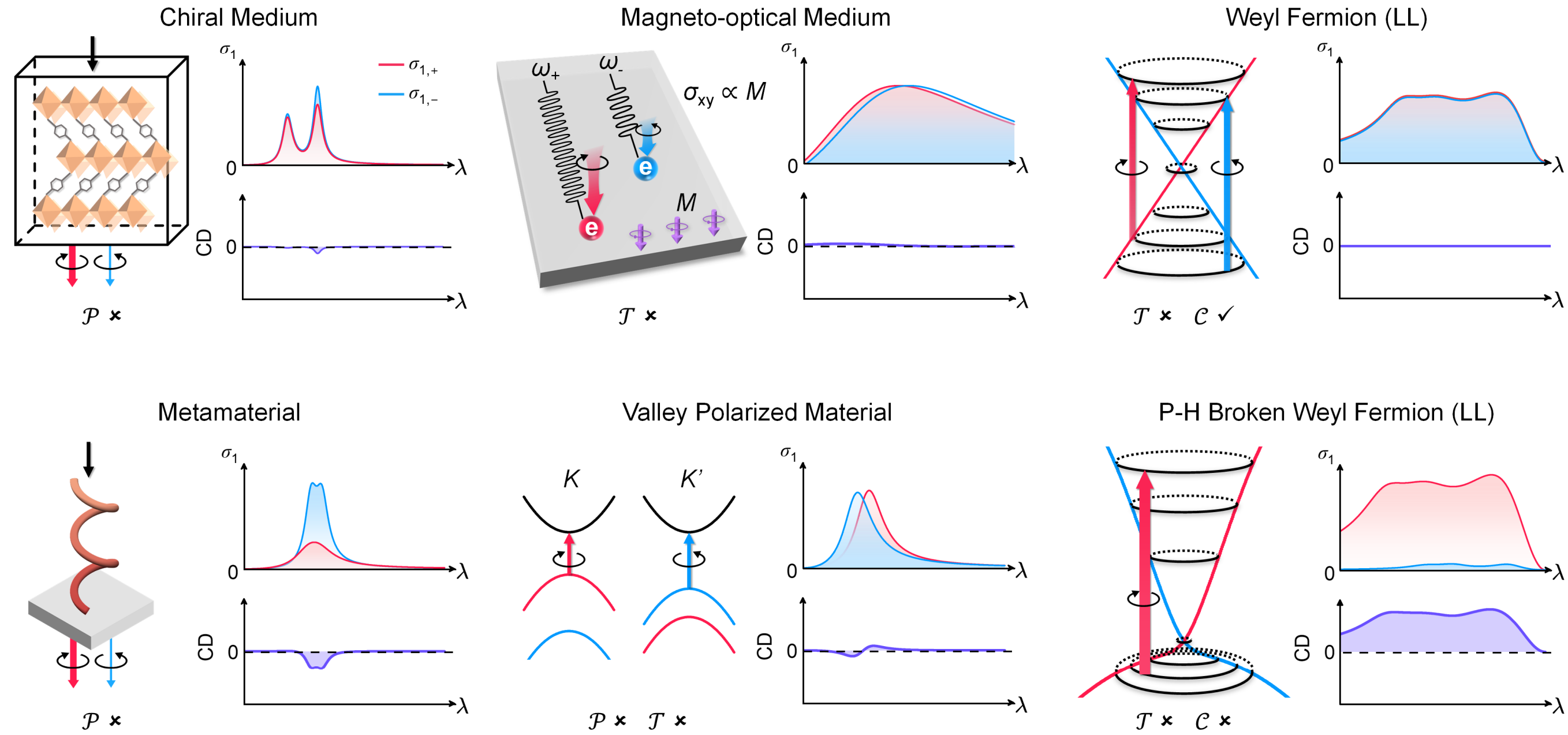

**Fig. 6 | Circular dichroism across representative material systems.** $\mathcal{P}$, $\mathcal{T}$, $\mathcal{C}$ denote parity, time-reversal, and particle-hole (P-H) symmetry, respectively. Organic molecules[33,34] and chiral crystals[35,36], where parity symmetry ($\mathcal{P}$) breaking dominates, typically exhibit weak circular dichroism within specific wavelengths. Metamaterials, including helicoid crystal arrays and gold-helix structures[37,38], enhance circular dichroism accompanied by resonance effects and requirement of precise structural tuning. Breaking time-reversal symmetry ($\mathcal{T}$) offers another pathway, as in magnetic materials[39], in which magneto-optical effects induce circular dichroism. In monolayer transition metal dichalcogenides, the combination of $\mathcal{P}$ breaking and strong spin-orbit coupling induces spin-splitting, thereby establishing the spin-valley locking paired by the time reversal symmetry[15]. Measurable circular dichroism emerges when $\mathcal{T}$ is additionally broken[40–42], where it further relies on extrinsic Pauli blocking and often changes sign with photon wavelength. Although ideal Weyl fermions are intrinsically chiral, their circular dichroism is strictly zero. Breaking particle-hole symmetry ($\mathcal{C}$) lifts this cancellation and gives rise to strong and broadband circular dichroism.